\documentclass[journal]{IEEEtran}

\usepackage{graphicx} 
\usepackage{subfigure}
\usepackage{amsmath}
\begin{document}

\title{\tiny\centering \vspace{-2mm}This work has been submitted to the IEEE Nuclear Science Symposium 2014 for publication in the conference record. Copyright may be transferred without notice, after which this version may no longer be available.\\
\Huge
Performance of a Large-Area\\
GEM Detector Prototype for the\\
Upgrade of the CMS Muon Endcap System}

\author{
D.~Abbaneo$^{15}$,
M.~Abbas$^{15}$,
M.~Abbrescia$^{2}$,
A.A.~Abdelalim$^{8}$,
M.~Abi~Akl$^{13}$,
W.~Ahmed$^{8}$,
W.~Ahmed$^{17}$,
P.~Altieri$^{2}$,
R.~Aly$^{8}$,
C.~Asawatangtrakuldee$^{3}$,
A.~Ashfaq$^{17}$,
P.~Aspell$^{15}$,
Y.~Assran$^{7}$,
I.~Awan$^{17}$,
S.~Bally$^{15}$,
Y.~Ban$^{3}$,
S.~Banerjee$^{19}$,
P.~Barria$^{5}$,
L.~Benussi$^{14}$,
V.~Bhopatkar$^{22\ast}$,~\IEEEmembership{Member,~IEEE,}
S.~Bianco$^{14}$,
J.~Bos$^{15}$,
O.~Bouhali$^{13}$,
S.~Braibant$^{4}$,
S.~Buontempo$^{24}$,
C.~Calabria$^{2}$,
M.~Caponero$^{14}$,
C.~Caputo$^{2}$,
F.~Cassese$^{24}$,
A.~Castaneda$^{13}$,
S.~Cauwenbergh$^{16}$,
F.R.~Cavallo$^{4}$,
A.~Celik$^{9}$,
M.~Choi$^{31}$,
K.~Choi$^{31}$,
S.~Choi$^{29}$,
J.~Christiansen$^{15}$,
A.~Cimmino$^{16}$,
S.~Colafranceschi$^{15}$,
A.~Colaleo$^{2}$,
A.~Conde~Garcia$^{15}$,
M.M.~Dabrowski$^{15}$,
G.~De~Lentdecker$^{5}$,
R.~De~Oliveira$^{15}$,
G.~de~Robertis$^{2}$,
S.~Dildick$^{9,16}$,
B.~Dorney$^{15}$,
W.~Elmetenawee$^{8}$,
G.~Fabrice$^{27}$, 
M.~Ferrini$^{14}$,
S.~Ferry$^{15}$,
P.~Giacomelli$^{4}$,
J.~Gilmore$^{9}$,
L.~Guiducci$^{4}$,
A.~Gutierrez$^{12}$,
R.M.~Hadjiiska$^{28}$,
A.~Hassan$^{8}$,
J.~Hauser$^{21}$,
K.~Hoepfner$^{1}$,
M.~Hohlmann$^{22\ast}$,~\IEEEmembership{Member,~IEEE,}
H.~Hoorani$^{17}$,
Y.G.~Jeng$^{18}$,
T.~Kamon$^{9}$,
P.E.~Karchin$^{12}$,
H.S.~Kim$^{18}$,
S.~Krutelyov$^{9}$,
A.~Kumar$^{11}$,
J.~Lee$^{31}$,
T.~Lenzi$^{5}$,
L.~Litov$^{28}$,
F.~Loddo$^{2}$,
T.~Maerschalk$^{5}$,
G.~Magazzu$^{26}$,
M.~Maggi$^{2}$,
Y.~Maghrbi$^{13}$,
A.~Magnani$^{25}$,
N.~Majumdar$^{19}$,
P.K.~Mal$^{6}$,
K.~Mandal$^{6}$,
A.~Marchioro$^{15}$,
A.~Marinov$^{15}$,
J.A.~Merlin$^{15}$,
A.K.~Mohanty$^{23}$,
A.~Mohapatra$^{22}$,
S.~Muhammad$^{17}$,
S.~Mukhopadhyay$^{19}$,
M.~Naimuddin$^{11}$,
S.~Nuzzo$^{2}$,
E.~Oliveri$^{15}$,
L.M.~Pant$^{23}$,
P.~Paolucci$^{24}$,
I.~Park$^{31}$,
G.~Passeggio$^{24}$,
B.~Pavlov$^{28}$,
B.~Philipps$^{1}$,
M.~Phipps$^{22}$,
D.~Piccolo$^{14}$,
H.~Postema$^{15}$,
G.~Pugliese$^{2}$,
A.~Puig Baranac$^{15}$,
A.~Radi$^{7}$,
R.~Radogna$^{2}$,
G.~Raffone$^{14}$,
S.~Ramkrishna$^{11}$,
A.~Ranieri$^{2}$,
C.~Riccardi$^{25}$,
A.~Rodrigues$^{15}$,
L.~Ropelewski$^{15}$,
S.~RoyChowdhury$^{19}$,
M.S.~Ryu$^{18}$,
G.~Ryu$^{31}$,
A.~Safonov$^{9}$,
A.~Sakharov$^{10}$,
S.~Salva$^{16}$,
G.~Saviano$^{14}$,
A.~Sharma$^{15}$,~\IEEEmembership{Senior Member,~IEEE,}
S.K.~Swain$^{6}$,
J.P.~Talvitie$^{15,20}$,
C.~Tamma$^{2}$,
A.~Tatarinov$^{9}$,
N.~Turini$^{26}$,
T.~Tuuva$^{20}$,
J.~Twigger$^{22}$,
M.~Tytgat$^{16}$,~\IEEEmembership{Member,~IEEE,}
I.~Vai$^{25}$,
M.~van~Stenis$^{15}$,
R.~Venditi$^{2}$,
E.~Verhagen$^{5}$,
P.~Verwilligen$^{2}$,
P.~Vitulo$^{25}$,
D.~Wang$^{3}$,
M.~Wang$^{3}$,
U.~Yang$^{30}$,
Y.~Yang$^{5}$,
R.~Yonamine$^{5}$,
N.~Zaganidis$^{16}$,
F.~Zenoni$^{5}$,
A.~Zhang$^{22}$%
\thanks{Manuscript received December 8, 2014.}%
\thanks{$^{1}$RWTH Aachen University, III Physikalisches Institut A, Aachen, Germany}%
\thanks{$^{2}$Politecnico di Bari, Universit\'{a} di Bari and INFN Sezione di Bari, Bari, Italy}
\thanks{$^{3}$Peking University, Beijing, China}%
\thanks{$^{4}$University and INFN Bologna, Bologna, Italy}%
\thanks{$^{5}$Universit\'{e} Libre de Bruxelles, Brussels, Belgium}%
\thanks{$^{6}$National Institute of Science Education and Research, Bhubaneswar, India}%
\thanks{$^{7}$Academy of Scientific Research and Technology, ENHEP, Cairo, Egypt}%
\thanks{$^{8}$Helwan University \& CTP, Cairo, Egypt}%
\thanks{$^{9}$Texas A\&M University, College Station, USA}%
\thanks{$^{10}$Kyungpook National University, Daegu, Korea}%
\thanks{$^{11}$University of Delhi, Delhi, India}%
\thanks{$^{12}$Wayne State University, Detroit, USA}%
\thanks{$^{13}$Texas A\&M University at Qatar, Doha, Qatar}%
\thanks{$^{14}$Laboratori Nazionali di Frascati - INFN, Frascati, Italy}%
\thanks{$^{15}$CERN, Geneva, Switzerland}%
\thanks{$^{16}$Ghent University, Dept. of Physics and Astronomy, Ghent, Belgium}%
\thanks{$^{17}$National Center for Physics, Quaid-i-Azam University Campus, Islamabad, Pakistan}%
\thanks{$^{18}$Chonbuk National University, Jeonju, Korea}%
\thanks{$^{19}$Saha Institute of Nuclear Physics, Kolkata, India}%
\thanks{$^{20}$Lappeenranta University of Technology, Lappeenranta, Finland}%
\thanks{$^{21}$University of California, Los Angeles, USA}%
\thanks{$^{22}$Florida Institute of Technology, Melbourne, USA}%
\thanks{$^{23}$Bhabha Atomic Research Centre, Mumbai, India}%
\thanks{$^{24}$INFN Napoli, Napoli, Italy}%
\thanks{$^{25}$INFN Pavia and University of Pavia, Pavia, Italy}%
\thanks{$^{26}$INFN Sezione di Pisa, Pisa, Italy}%
\thanks{$^{27}$IRFU CEA-Saclay, Saclay, France}%
\thanks{$^{28}$Sofia University, Sofia, Bulgaria}%
\thanks{$^{29}$Korea University, Seoul, Korea}%
\thanks{$^{30}$Seoul National University, Seoul, Korea}%
\thanks{$^{31}$University of Seoul, Seoul, Korea}%
\thanks{$^{\ast}$Corresponding authors: vbhopatkar2010@my.fit.edu, hohlmann@fit.edu}
}

\maketitle
\pagestyle{plain}

\begin{abstract}
Gas Electron Multiplier (GEM) technology is being considered for the forward muon upgrade of the CMS experiment in Phase 2 of the CERN LHC. Its first implementation is planned for the GE1/1 system in the $1.5 < \mid\eta\mid < 2.2$ region of the muon endcap  mainly to control muon level-1 trigger rates after the second long LHC shutdown. A GE1/1 triple-GEM detector is read out by 3,072 radial strips with 455 $\mu$rad pitch arranged in eight $\eta$-sectors. We assembled a full-size GE1/1 prototype of 1m length at Florida Tech and tested it in 20-120 GeV hadron beams at Fermilab using Ar/CO$_{2}$ 70:30 and the RD51 scalable readout system. Four small GEM detectors with 2-D readout and an average measured azimuthal resolution of 36 $\mu$rad provided precise reference tracks. Construction of this largest GEM detector built to-date is described. Strip cluster parameters, detection efficiency, and spatial resolution are studied with position and high voltage scans. The plateau detection efficiency is \mbox{[97.1 $\pm$ 0.2 (stat)]\%.} The azimuthal resolution is found to be \mbox{[123.5 $\pm$ 1.6 (stat)] $\mu$rad} when operating in the center of the efficiency plateau and using full pulse height information. The resolution can be slightly improved by $\sim$ 10 $\mu$rad when correcting for the bias due to discrete readout strips. The CMS upgrade design calls for readout electronics with binary hit output. When strip clusters are formed correspondingly without charge-weighting and with fixed hit thresholds, a position resolution of \mbox{[136.8 $\pm$ 2.5 stat] $\mu$rad} is measured, consistent with the expected resolution of strip-pitch/$\sqrt{12}$ = 131.3 $\mu$rad. Other $\eta$-sectors of the detector show similar response and performance.
\end{abstract}

\section{Introduction}
\IEEEPARstart{D}{uring} the phase II upgrade of the Compact Muon Solenoid (CMS) experiment at the CERN, CMS GEM collaboration intends to install large-area GEM \cite{GEM} detectors in the forward muon endcap in the high-$\eta$ region \mbox{$1.5 < \mid\eta\mid < 2.2$.} Fig. 1 shows the quadrant of the muon system, where installation of GE1/1 detectors is proposed. This installation will help to restore redundancy for tracking and triggering in the muon system, as GEM detectors provide very precise tracking information due to high spatial resolution. They can also sustain high particles rates up to $\sim$MHz/cm$^{2}$. Cathode Strips chambers (CSC) alone many times misidentify multiply scattered lower p$_{T}$ muons as high p$_{T}$ muons. This problem can be overcome by using GEM detectors in conjuction with the CSC system as shown in Fig. 2; together they provide an accurate measurement of the muon bending angle that is not affected by multiple scattering. This discriminates lower p$_{T}$ muons from higher p$_{T}$ muons and reduces the soft muon rate at the level-1 trigger as shown in Fig. 3, which will help control the muon trigger rate at the high luminosity LHC. 
\begin{figure}[!hb]
\centering
\includegraphics[width=0.5\textwidth]{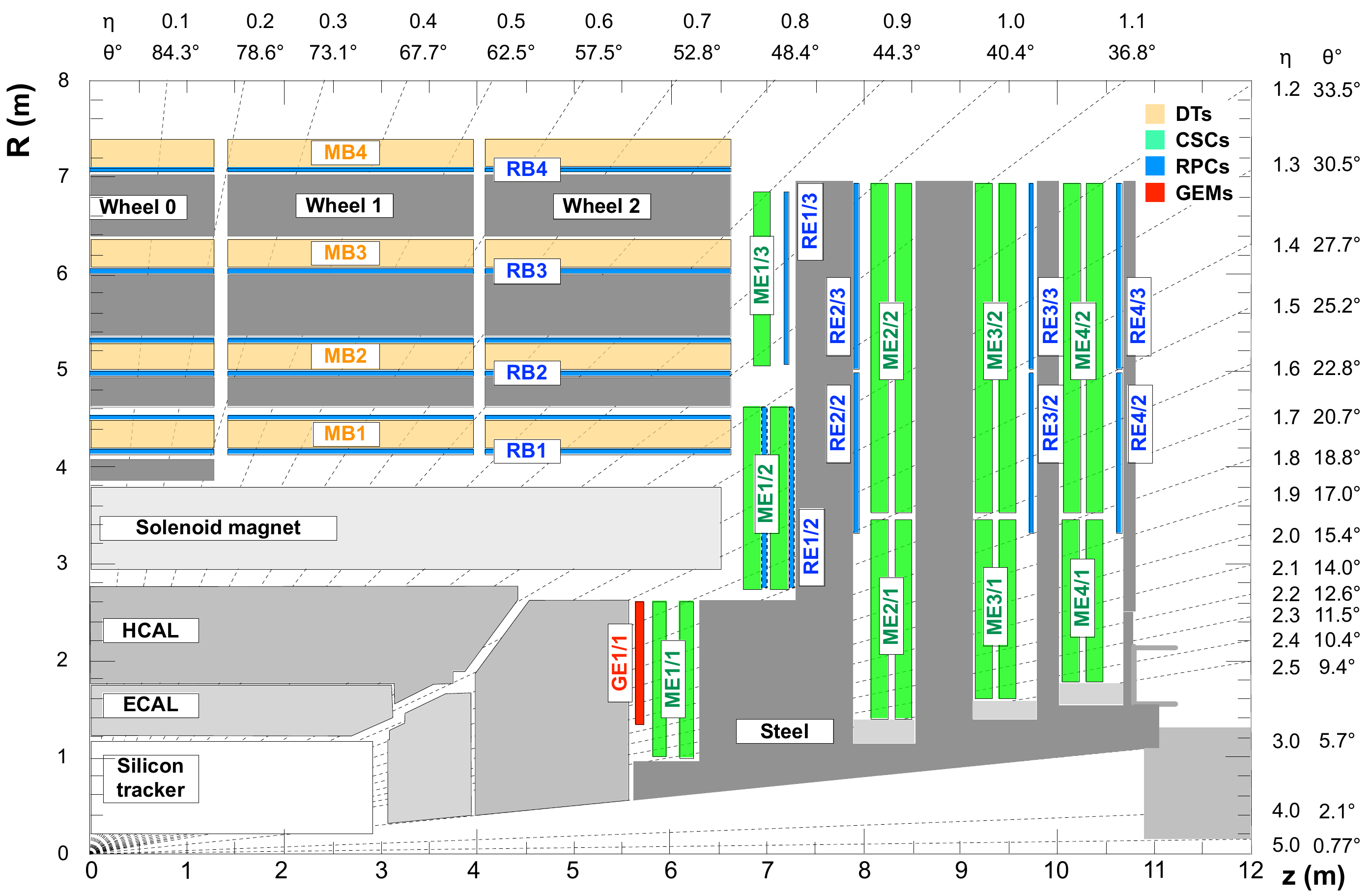}
\caption{A quadrant of the CMS muon system with proposed GE1/1 upgrade in red.}
\label{label1}
\end{figure}
\begin{figure}[!hb]
\centering
\includegraphics[width=0.4\textwidth, height = 1.75in]{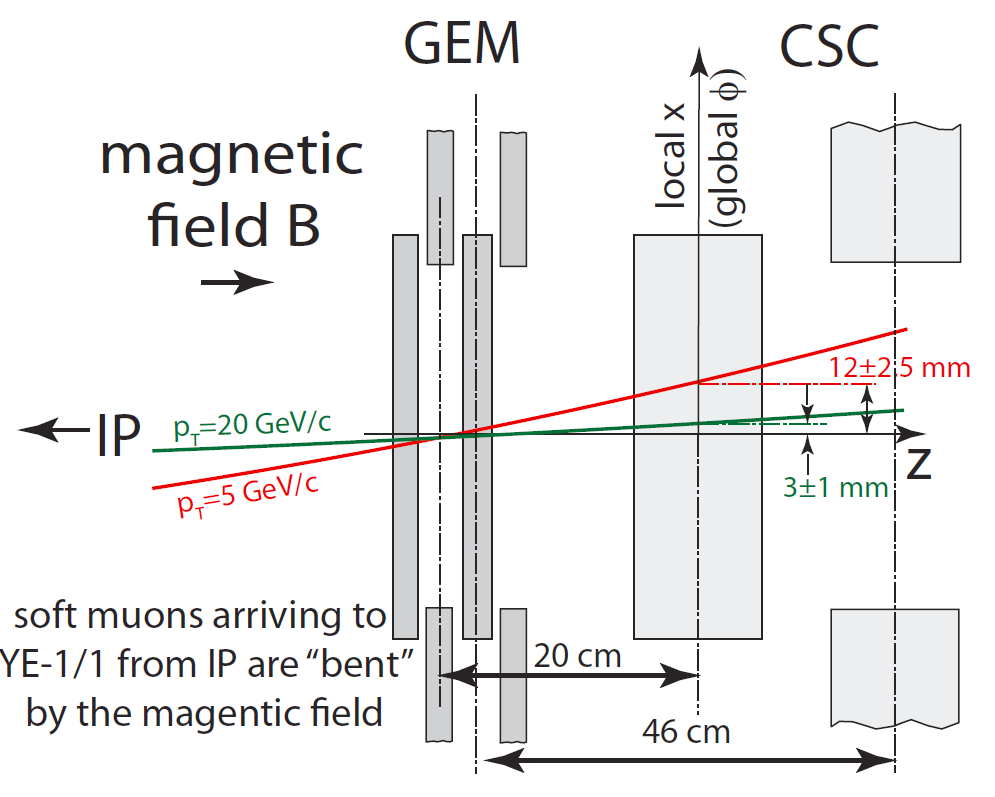}
\caption{GEM and CSC system in each station enlarge the lever arm for a bending angle measurement unaffected by multiple scattering.}
\label{label25}
\end{figure}

\begin{figure}[!htb]
\centering
\subfigure[]{\label{label2-2}}
\centering
\includegraphics[width=0.46\textwidth]{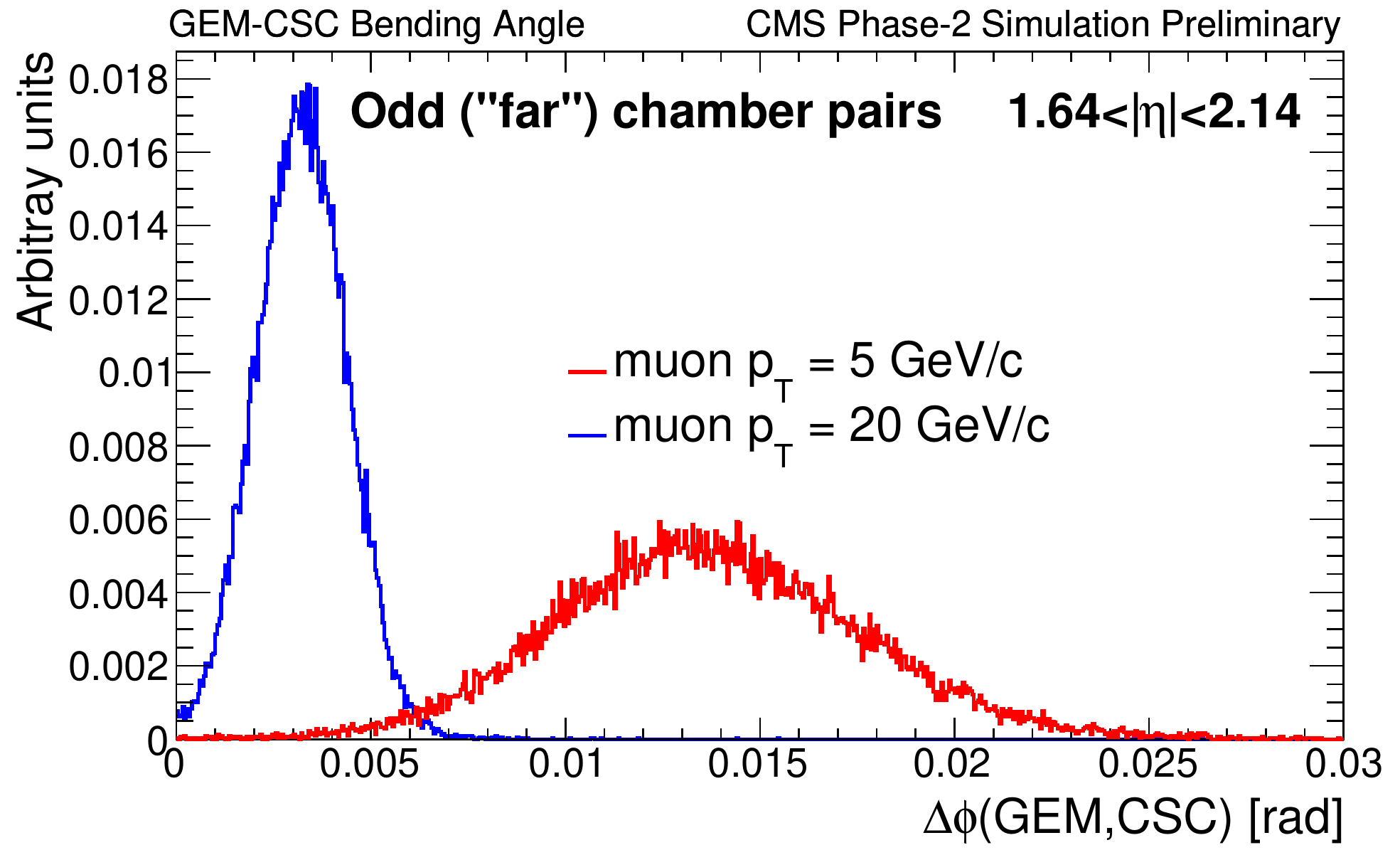}
\newline
\subfigure[]{\label{label2-3}}
\centering
\includegraphics[width=0.46\textwidth, height= 2.5in]{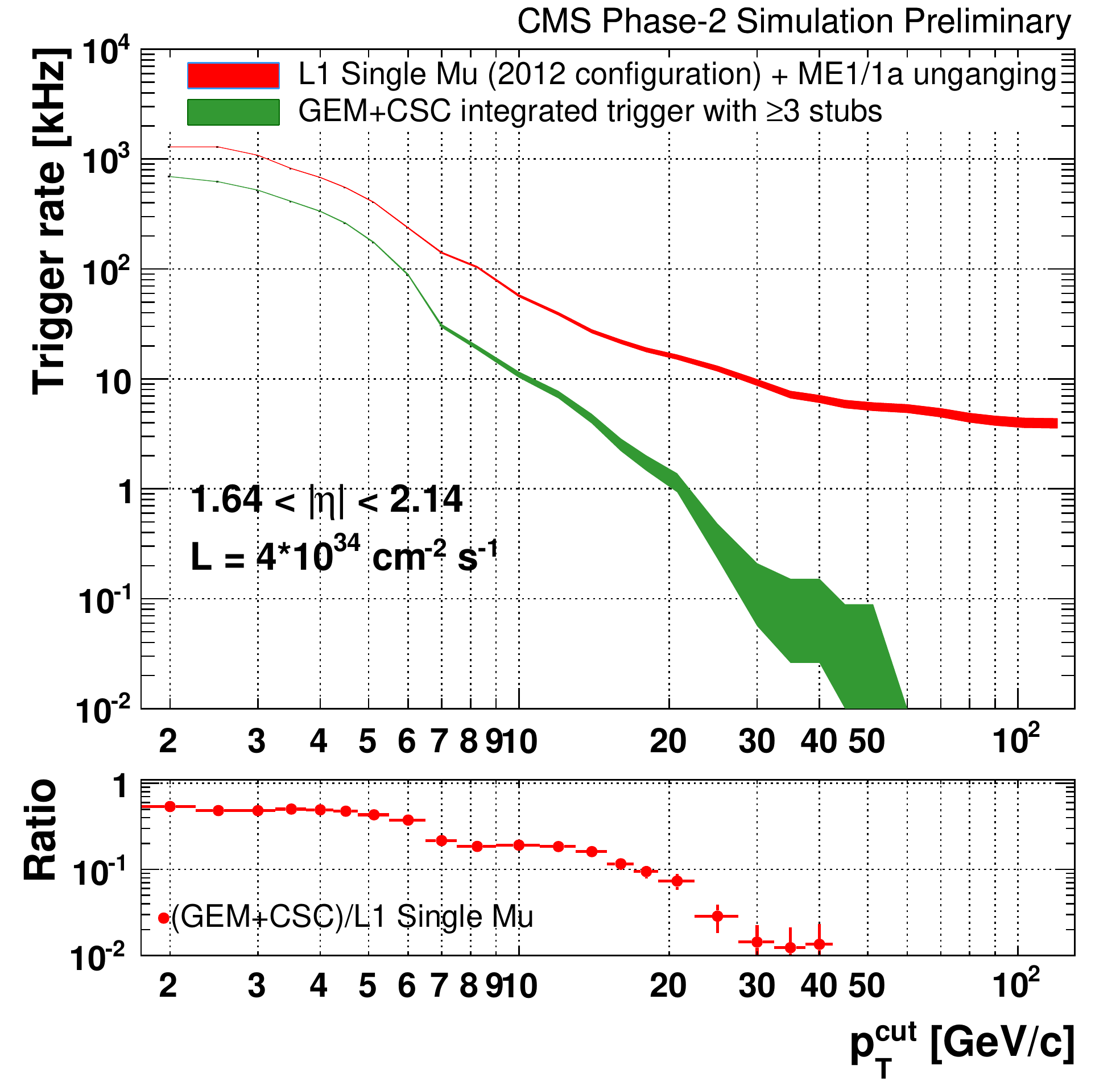}
\caption{(a) Simulation of the bending angle measurement in the first endcap station (GE1/1-ME1/1) for soft ($\sim$5 GeV) and hard ($\sim$20 GeV) muons \cite{Michael}. (b) Simulation of the inclusive muon trigger rate expected for the LHC Phase 2 as function of the Level-1 p$_{T}$ trigger threshold in \mbox{$1.5 < \mid\eta\mid < 2.2$.}}
\label{label2}
\end{figure}
\newpage
We have constructed a 1m-long GE1/1 prototype, one of the largest GEM detectors in existance, at Florida Institute of Technology and tested it in a hadron beam at Fermilab. This particular prototype represents the third generation of GE1/1 prototypes. Performance characteristics of the detector such as strip cluster parameters, detection efficiency, and spatial resolution have been studied using full pulse height information as well as with binary hit reconstruction. A correction for the non-linear strip response is also implemented in an attempt to improve overall spatial resolution of the GE1/1 detector.

\section{Construction of a Large-Area GEM detector at Florida Tech}

The large-area GEM detector proposed for the CMS muon endcap upgrade is a trapezoidal Triple-GEM detector with approximately 99$\times$(28-45) cm$^{2}$ active surface area. This detector has a 3/1/2/1 mm (drift, transfer 1, transfer 2, induction) internal electrode gap configuration. The GEM foils used in this detector are produced by single-mask etching technique at CERN. Each GEM foil is divided into 35 high voltage sectors that are transverse to the direction of the readout strips. A stack of these three GEM foils is mounted on the drift electrode; the gas volume is closed with the readout board. This detector is constructed using an internal mechanical stretching method \cite{IEEE2012} introduced in 2011. There are a total of 3072 radial readout strips with 455 $\mu$rad pitch along the azimuthal direction and distributed over eight $\eta$-sectors. In each of the sectors, induced signals are read out via 384 radial strips. Fig. 4 shows the main steps involved in the construction of the GE1/1-III prototype detector.

\begin{figure}[h]
\centering
\subfigure[]{\label{label3-1}}
\centering
\includegraphics[width=0.48\textwidth]{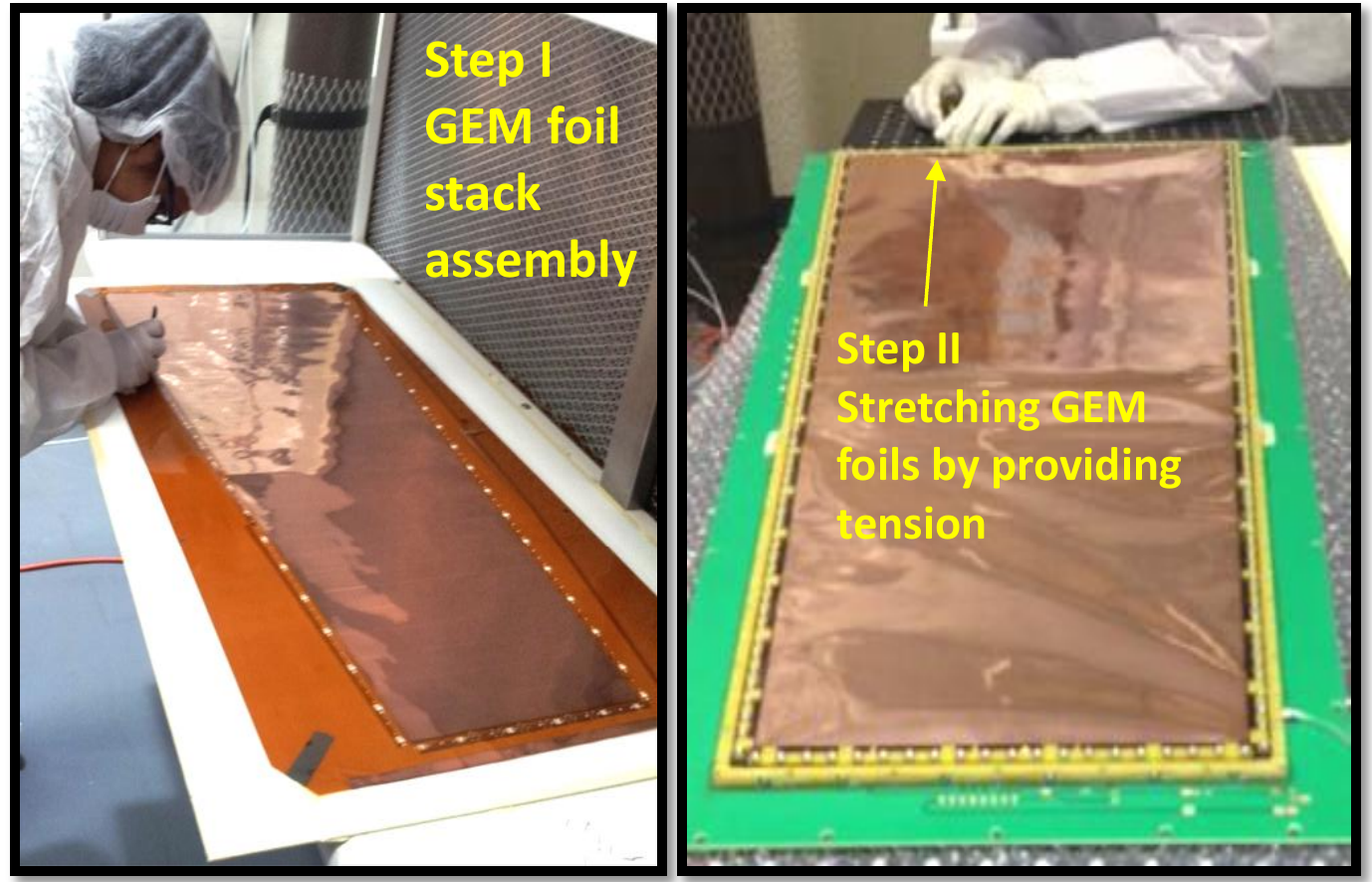}
\subfigure[]{\label{label3-3}}
\centering
\includegraphics[width=0.48\textwidth]{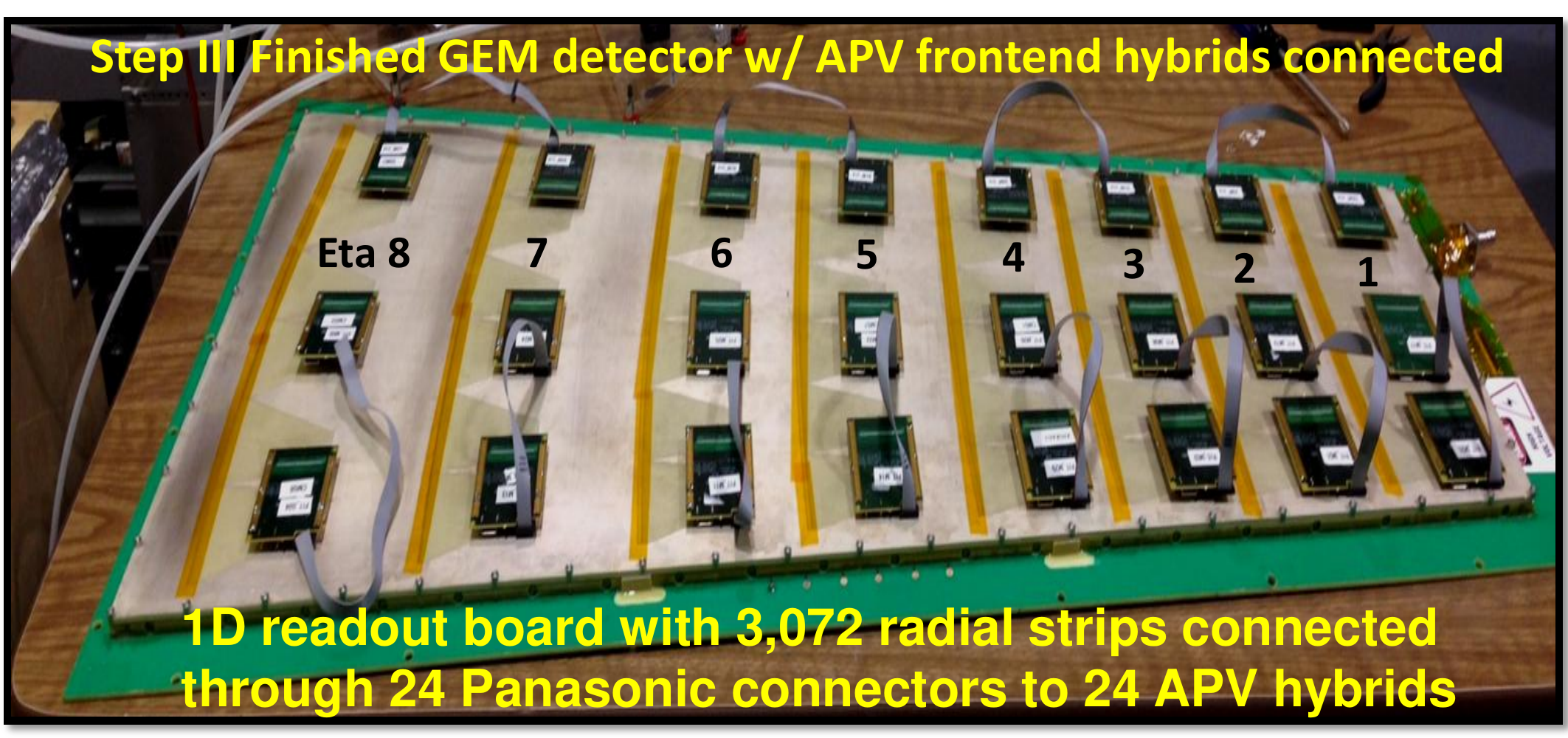}
\caption{Construction of the 1m-long large-area GE1/1-III prototype in three steps. The instrumentation with APV front-end hybrids and the numbering of eta sectors is shown in (b).}
\label{label3}
\end{figure}

\section{Test Beam setup}

The GE1/1-III prototype detector was tested in a \mbox{32 GeV} mixed hadron beam at the Fermilab test beam facility (FTBF) in October 2013. For tracking studies, this detector was positioned on a movable table between four GEM detectors with 2D Cartesian readout \cite{Ketzer}. Three of them were small \mbox{10 cm $\times$ 10 cm} GEM detectors and one was a \mbox{50 cm $\times$ 50 cm} (Provided by University of Virginia) GEM detector for which only the central 10 cm $\times$ 10 cm area was instrumented shown in Fig. 5. These 2D GEM detectors contained 256 straight strips along each horizontal (y-coordinate) and \mbox{vertical (x-coordinate)} plane with 0.4 mm pitch. The data were collected using the RD51 scalable readout system \cite{IEEE2010} \cite{Amore} with an external trigger provided by scintillators. During this beam test, all detectors used an Ar/CO$_{2}$ 70:30 gas mixture.

\begin{figure}[!h]
\centering
\includegraphics[width=4.7in]{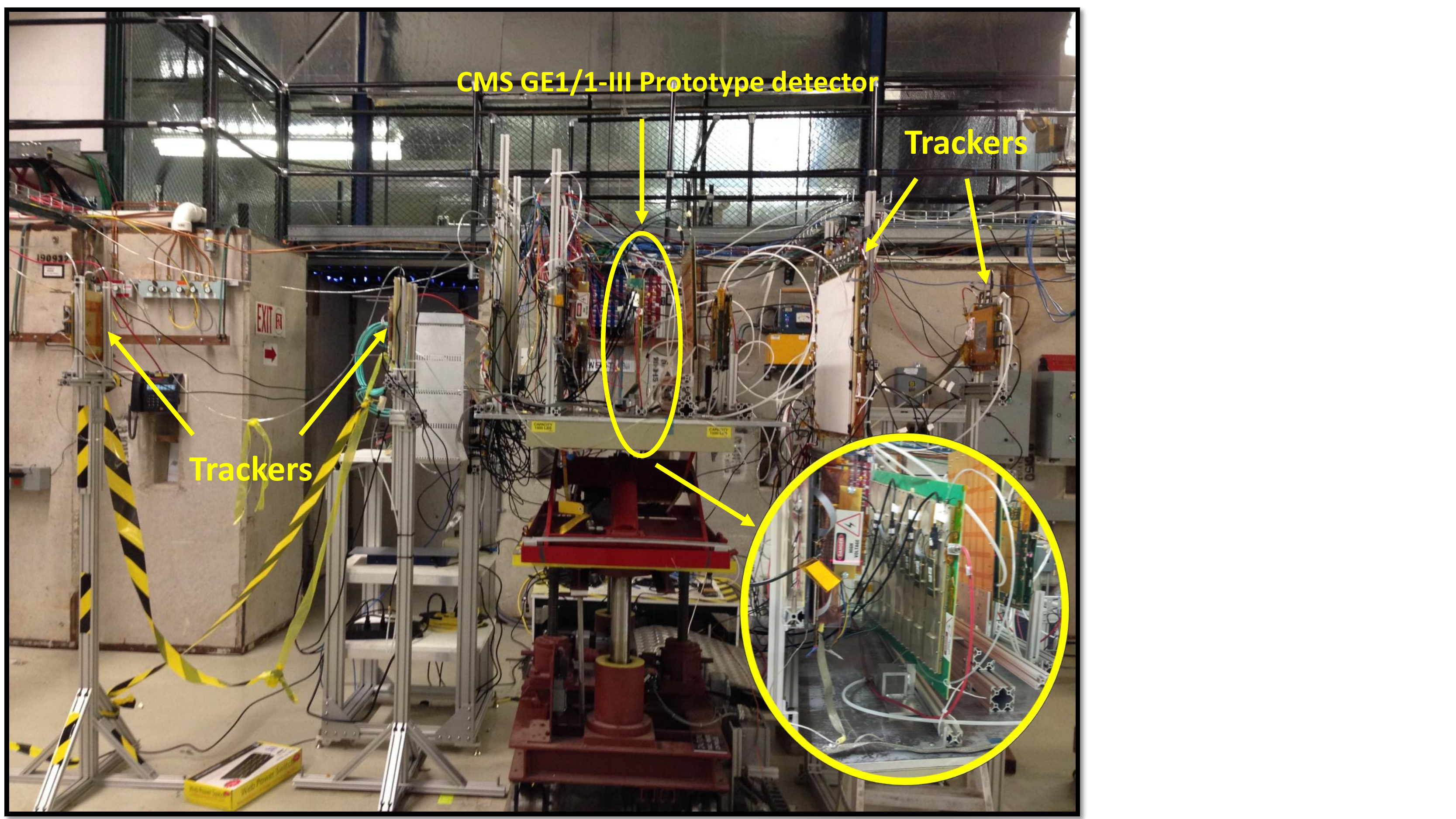}
\caption{Experimental setup in the beam line MT6.2-B at FTBF.}
\label{fig_4}
\end{figure}

\section{Beam Test results}

\subsection{Performance Characteristics}
Performance characteristics of the GE1/1 detector were studied using high voltage and position scan data. In the high voltage scan, the beam was focused on sector 5 and the voltage of the GE1/1 detector was varied from 2900V to 3350V, whereas voltages of all tracker detectors were set so that the trackers operated on their efficiency plateaus throughout all measurements. Fig. 6 shows the cluster charge distribution for $\eta$-sector 5 at 3250V. All distributions obtained at different high voltages are fitted with a Landau function and the Most Probable Values (MPV) obtained from these fits are used for obtaining the uniformity results for the GE1/1 detector.\\
\begin{figure}[!htb]
\centering
\includegraphics[width=0.55\textwidth]{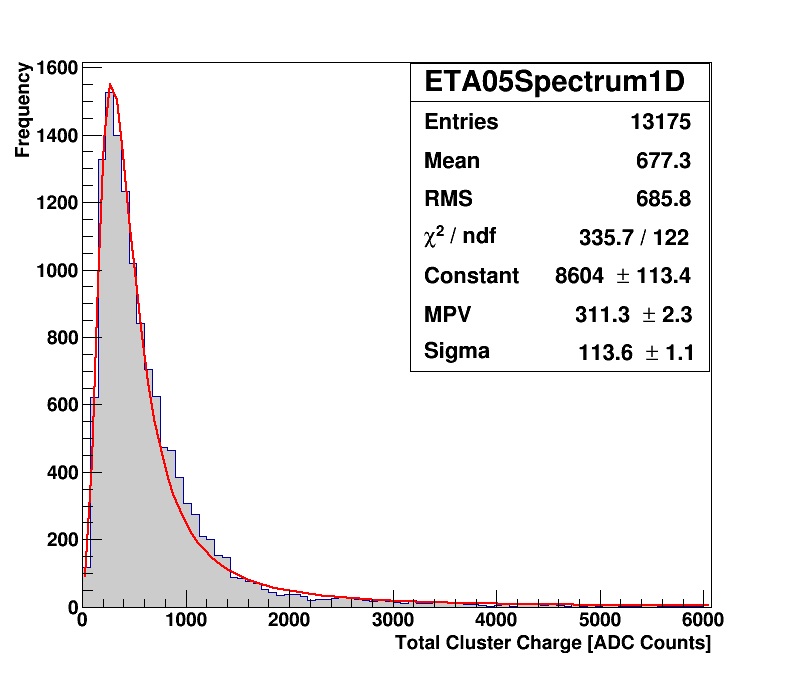}
\caption{GE1/1 cluster charge distribution at 3250V fitted with Landau function.}
\label{label6}
\end{figure}
\pagebreak

The number of strips above threshold in a strip cluster define the strip multiplicity of that cluster. For the GE1/1 detector, the distribution of multiplicity in strip clusters is shown in \mbox{Fig. 7;} the average strip multiplicity is 2.4 strips. Fig. 8 shows that the strip multiplicity increases with high voltage, i.e. gas gain.

\begin{figure}[!h]
\centering
\includegraphics[width=0.5\textwidth]{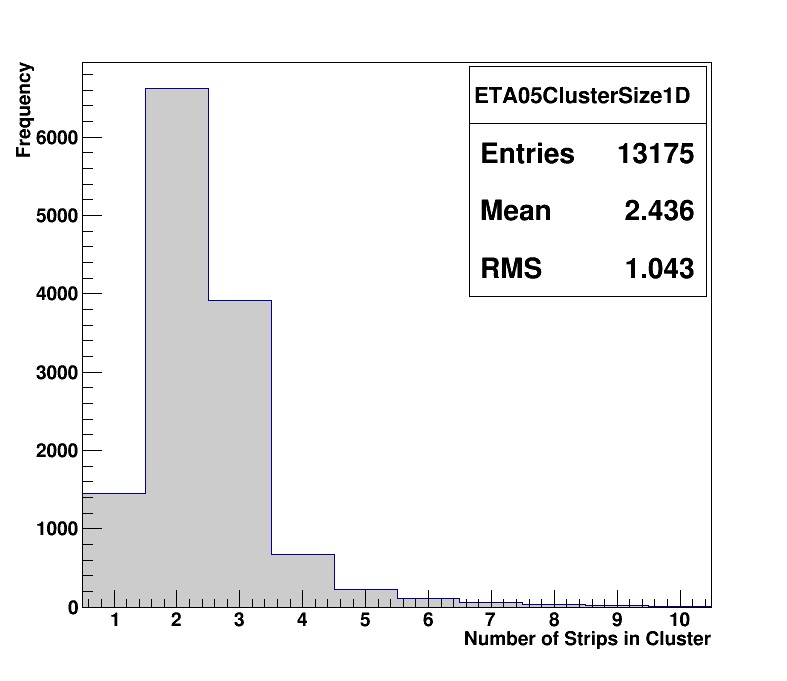}
\caption{Strip multiplicity in GE1/1 clusters at 3250V.}
\label{label7}
\end{figure}
\begin{figure}[!htb]
\centering
\includegraphics[width=0.5\textwidth, height=3in]{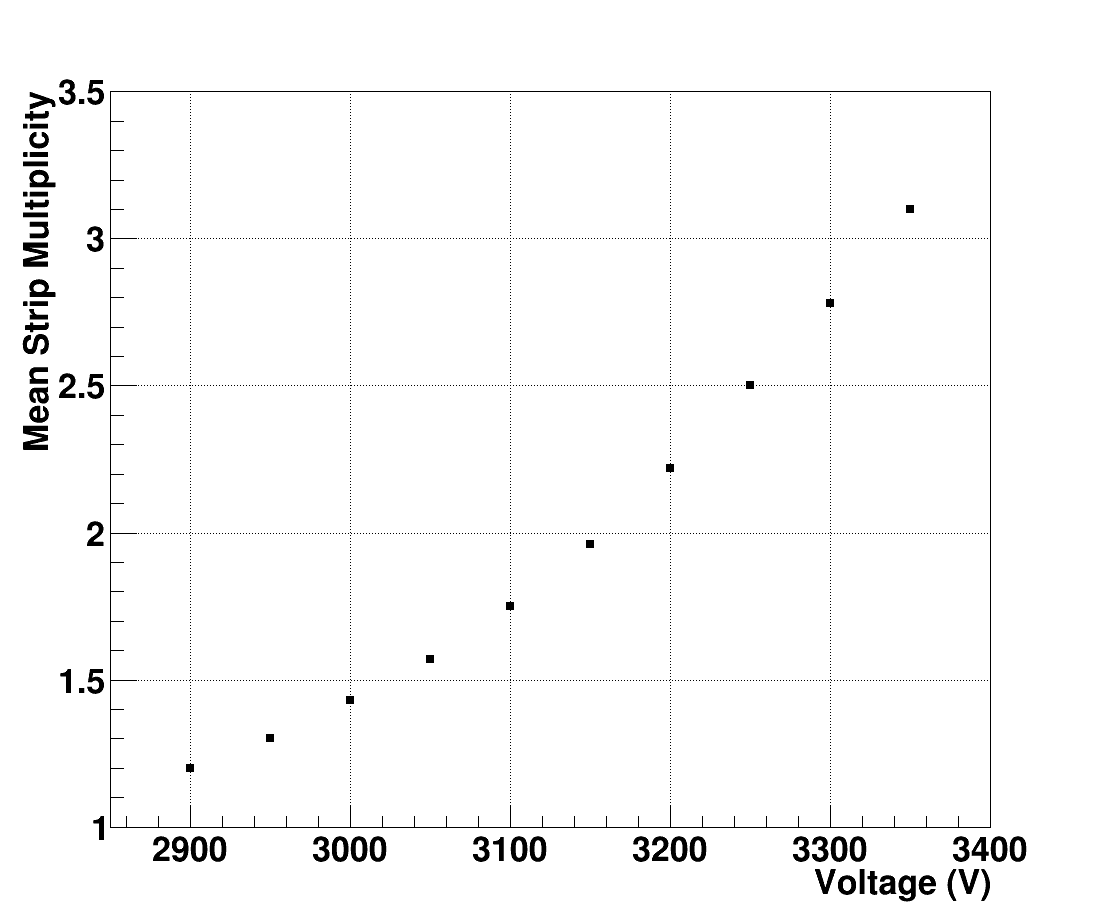}
\caption{Mean strip multiplicity increases with high voltage.}
\label{label8}
\end{figure}

The detection efficiency is obtained as 
$\epsilon = \frac{N_{1}}{N-N_{2}}$,
where \mbox{N = total number of triggered events;}
$N_{1}$ = number of events with Cluster Multiplicity \mbox{(CM) $\geq$ 1 for given sector;}
\mbox{$N_{2}$ = number of events with CM $\geq$ 1 for other sectors.}
A few strips from neighboring sectors triggered sometimes due to scattering of particles or because the beam was hitting near the edge of the sector. To obtain accurate efficiency of the given sector, we eliminate such events from the total number of events. The overall detection efficiency of the detector is measured for four threshold cuts on the pedestal width, namely 3$\sigma$, 4$\sigma$, 5$\sigma$, and 6$\sigma$, where $\sigma$ is the width of the pedestal distribution from a Gaussian fit. Fig. 9 shows efficiency curves fitted with a sigmoid function for these threshold cuts. The efficiency curves show a long plateau for higher voltages and the efficiency values on plateau are not much affected by different threshold cuts. The detection efficiency of the GE1/1 detector with 5$\sigma$ cut on the pedestal width is \mbox{[97.1 $\pm$ 0.2 (stat)]\%.}
\begin{figure}[!htb]
\centering
\includegraphics[width=0.5\textwidth]{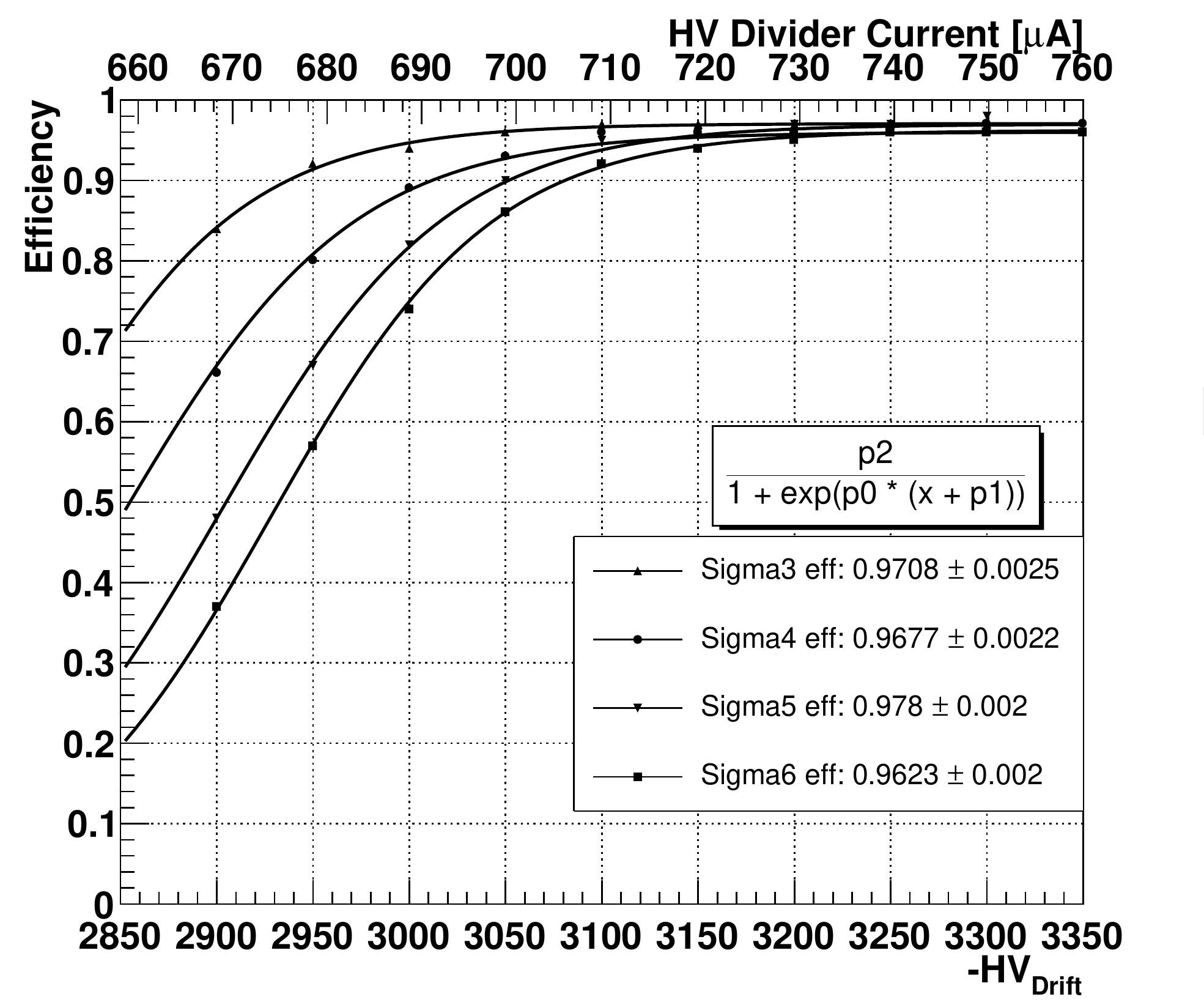}
\caption{Detection efficiency with different strip threshold cut on pedestal width.}
\label{label9}
\end{figure}
\begin{figure}[!hbt]
\centering
\includegraphics[width=0.5\textwidth]{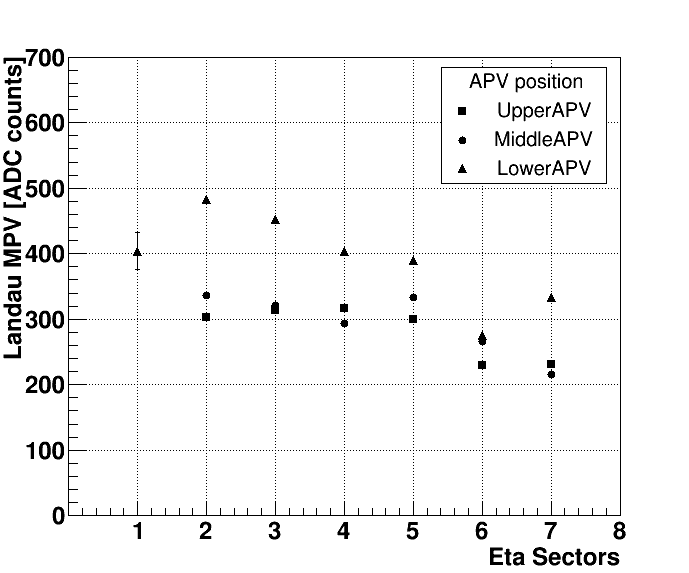}
\caption{Response uniformity in seven $\eta$-sectors for three different APV positions, i.e. Upper, Middle, and Lower at 3250V.}
\label{label10}
\end{figure}

The position scan is used to measure the uniformity response of the detector for all $\eta$-sectors at operating voltage 3250V. The scan was performed for three APV positions (see Fig. 4 (b)). For each sector, the charge uniformity was measured using the Landau MPV from the cluster charge distribution. Fig. 10 shows the uniformity response for the first seven sectors. The uniformity varies by $\sim$25\% across different sectors. The variation in sector 6 and 7 has been improved in subsequent prototypes by carefully stretching foils in that region and by minimizing any bending of pcbs.

The response uniformity of the detector is also calculated by considering only single-strip single-cluster events. Fig. 11 shows the uniformity response of the $\eta$-sectors of the detector.
\begin{figure}[!htb]
\centering
\includegraphics[width=0.5\textwidth]{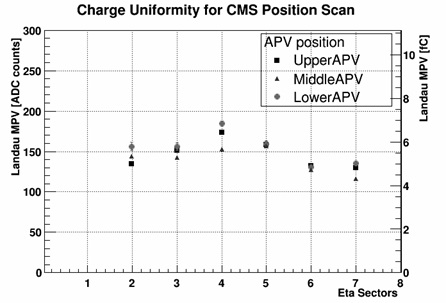}
\caption{Charge uniformity for different $\eta$ sectors at 3250V using single-strip single-cluster events.}
\label{label17}
\end{figure}

\subsection{Tracking Analysis}
Beam profiles can be obtained using 2D hit maps of the reference tracker detectors. The beam profile for the secondary 32 GeV mixed-hadron beam is oval while the beam profile for the 120 GeV proton beam is more circular and more focused as shown in Fig. 12.
\begin{figure}[h]
\centering
\subfigure[]{\label{label5-1}}
\centering
\includegraphics[width=0.235\textwidth]{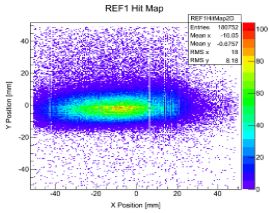}
\subfigure[]{\label{label5-2}}
\centering
\includegraphics[width=0.235\textwidth]{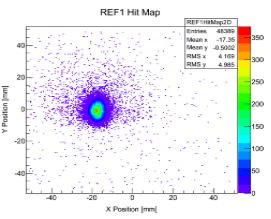}
\caption{Beam profiles: (a) 32 GeV hadron beam. (b) 120 GeV proton beam.}
\label{label5}
\end{figure}

The tracking analysis is done in three steps to study the spatial resolution of the GE1/1 detector. The first step is alignment which is performed in two steps. The first one is to shift all tracker detectors so that their origins in x-y coordinate coincide and their residuals are centered at zero. In the second step, the three downstream tracking detectors are rotated with respect to the first tracker detector by using initial shift parameters from step one. The rotating angle for each detector is optimized such that the residual width of each detector is minimized.

The next step in the tracking analysis is a transformation from Cartesian (x, y) coordinates to \mbox{polar (r, $\phi$) coordinates}, because the GE1/1 detector measures the \mbox{azimuthal coordinate ($\phi$)} with its radial strips. Fig. 13 shows correlation in $\phi$ for hits in the GE1/1 detector and in the first tracker detector.
\begin{figure}[!h]
\centering
\includegraphics[width=0.5\textwidth, height=2.5in]{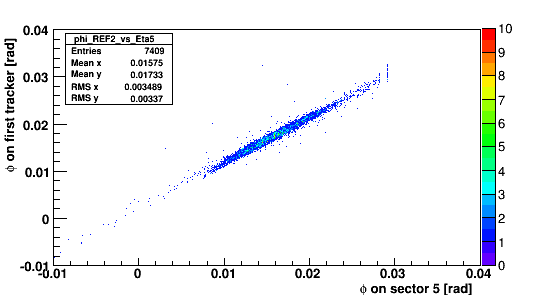}
\caption{Correlation of GE1/1 detector hits with hits in first tracker detector.}
\label{label11}
\end{figure}
\begin{figure}[!htb]
\centering
\subfigure[]{\label{label12-1}}
\centering
\includegraphics[width=0.5\textwidth, height=2.1in]{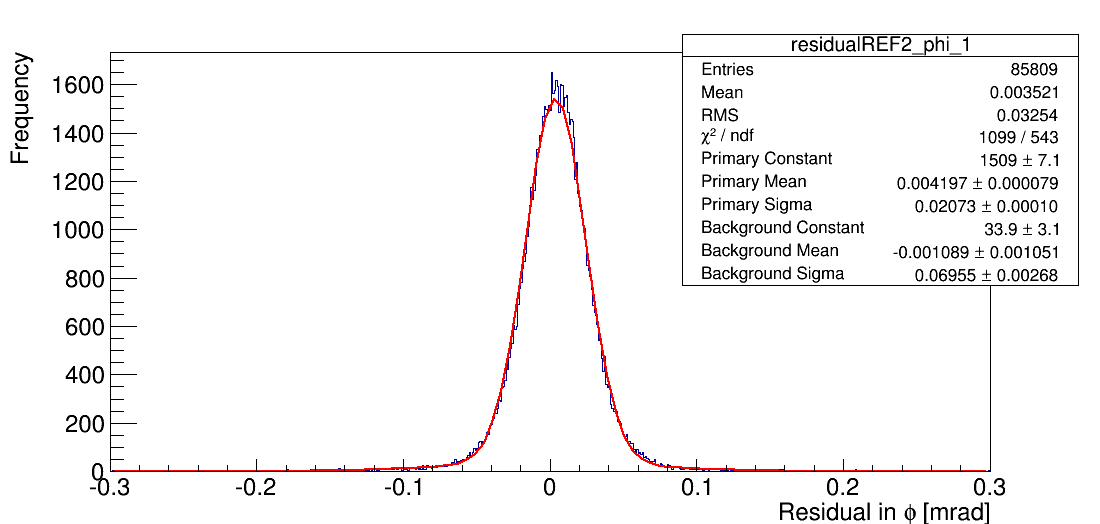}
\subfigure[]{\label{label12-2}}
\centering
\includegraphics[width=0.5\textwidth, height=2.1in]{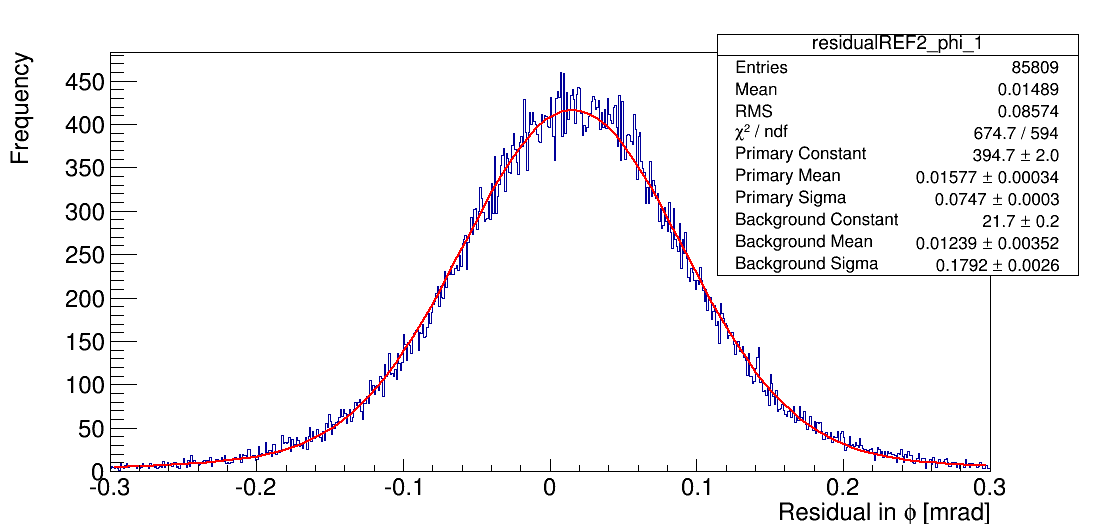}
\caption{Residuals for tracker 1 in $\varphi$ : (a) Inclusive residual with width \mbox{$\sigma$ = 21 $\mu$rad} and (b) Exclusive residual with width $\sigma$ = 75 $\mu$rad.}
\label{label12}
\end{figure}
\begin{figure}[!htb]
\centering
\includegraphics[width=0.5\textwidth, height=2.5in]{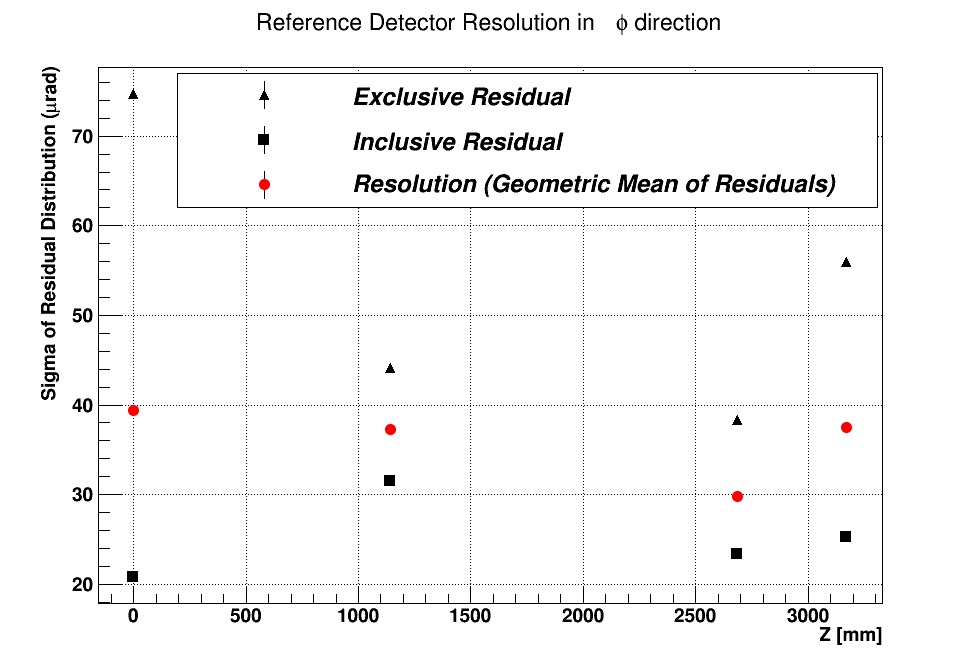}
\caption{Residuals and angular resolution of the tracking detectors}
\label{label13}
\end{figure}

In the final step of the tracking analysis, both inclusive and exclusive track-hit residuals are calculated. The definition of an inclusive (exclusive) residual is the residual calculated by including (excluding) the probed detector in the track fitting. The spatial resolution of the detector is calculated from the geometric mean \cite{Carnegie} \cite {Alex} of the widths of inclusive and exclusive residuals, i.e. $\sigma = \sqrt{\sigma_{inc}\times\sigma_{exc}}$ \cite{Arogancia}. Fig. 14 shows both residual widths for the first tracker detector. The inclusive residual width is \mbox{$\sigma_{inc}$ = 21 $\mu$rad} and exclusive residual width is \mbox{$\sigma_{exc}$ = 75 $\mu$rad;} with a geometric mean of $\sim$40 $\mu$rad. Similarly, the residual widths of the other three tracker detectors are calculated. Fig. 15 summarizes the angular resolution of all four tracking detectors.

The angular resolution of the GE1/1-III prototype GEM detector is calculated using two different methods for obtaining the hit position. The barycentric method uses the full pulse height information to find the strip cluster barycenter and the binary method uses binary hits reconstructed offline from the analog APV readout to emulate the behavior of VFAT \cite{VFAT} chips with binary output. The results are as follows:\\

\subsubsection{The Barycentric Method}
Fig. 16 shows inclusive and exclusive residuals of the GE1/1 detector at the center of sector 5 at 3250V using a 5$\sigma$ cut on the pedestal width. The inclusive residual is 110.7 $\mu$rad and the exclusive residual is 137.9 $\mu$rad in azimuthal direction; this corresponds to \mbox{208.1 $\mu$m} and 259.3 $\mu$m, respectively, with \mbox{R = 1880.5 mm.} The resulting resolution of the GE1/1 detector using pulse height analog readout is [123.3 $\pm$ 1.6 (stat)] $\mu$rad, i.e. 27\% of the strip pitch. 
\begin{figure}[!htb]
\centering
\subfigure[]{\label{label14-1}}
\centering
\includegraphics[width=0.5\textwidth, height=2.25in]{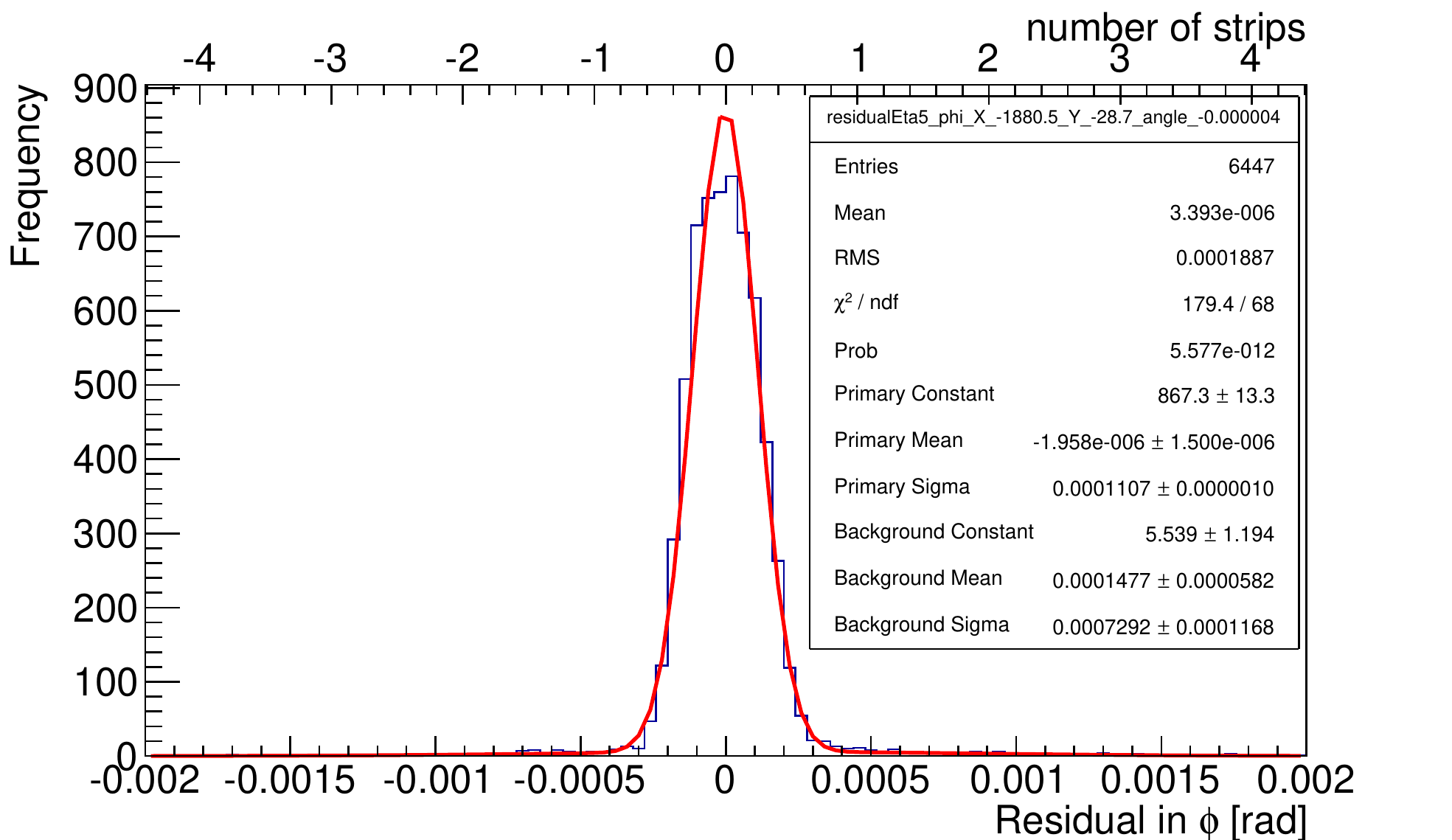}
\subfigure[]{\label{label14-2}}
\centering
\includegraphics[width=0.5\textwidth, height=2.25in]{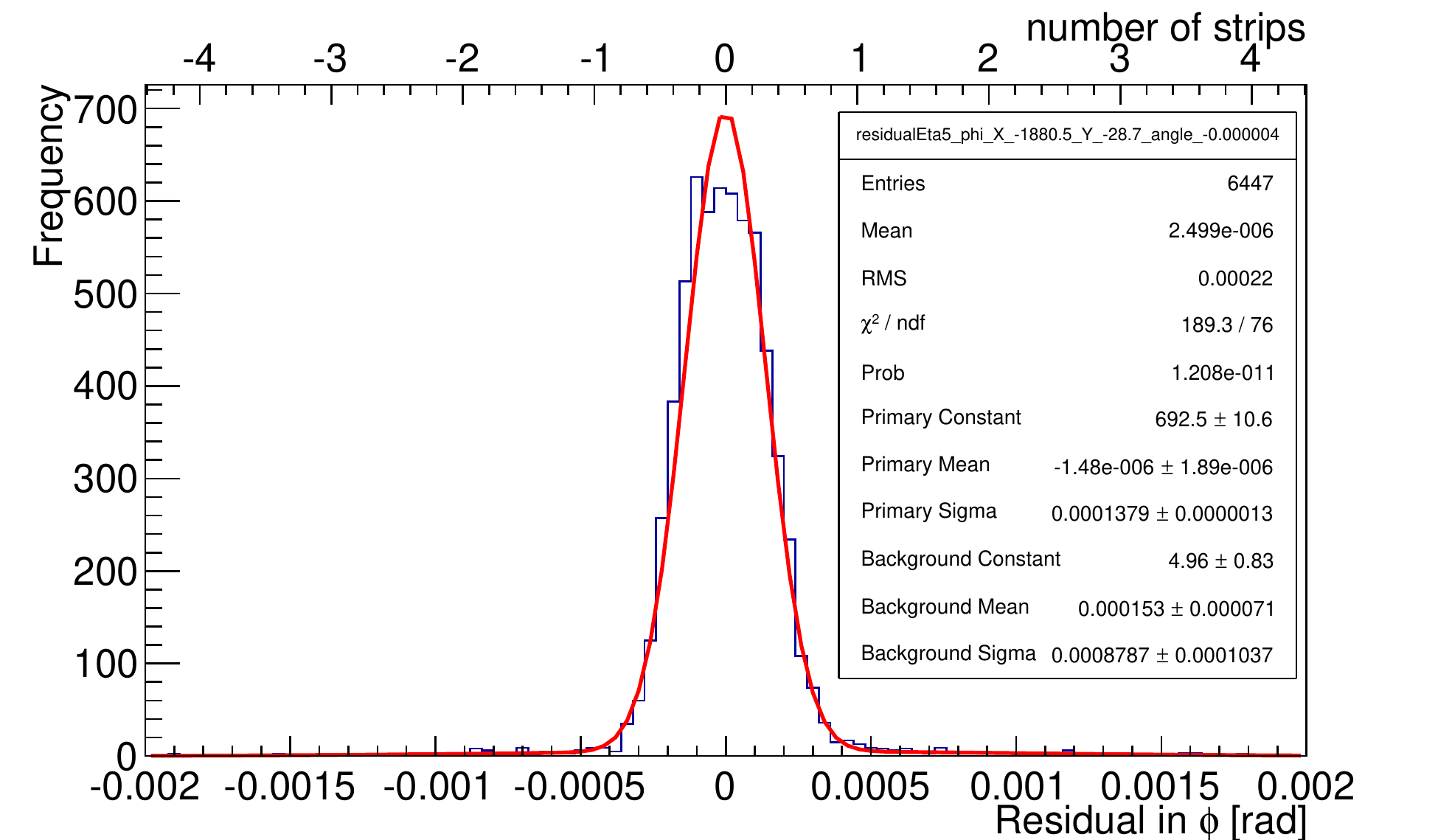}
\caption{(a) Inclusive GE1/1 residual with $\sigma$ = 110.7 $\mu$rad and (b) Exclusive GE1/1 residual with $\sigma$ = 137.9 $\mu$rad using full pulse height information and double-Gaussian fits.}
\label{label14}
\end{figure}
\begin{figure}[!htb]
\centering
\includegraphics[width=0.5\textwidth, height=2.2in]{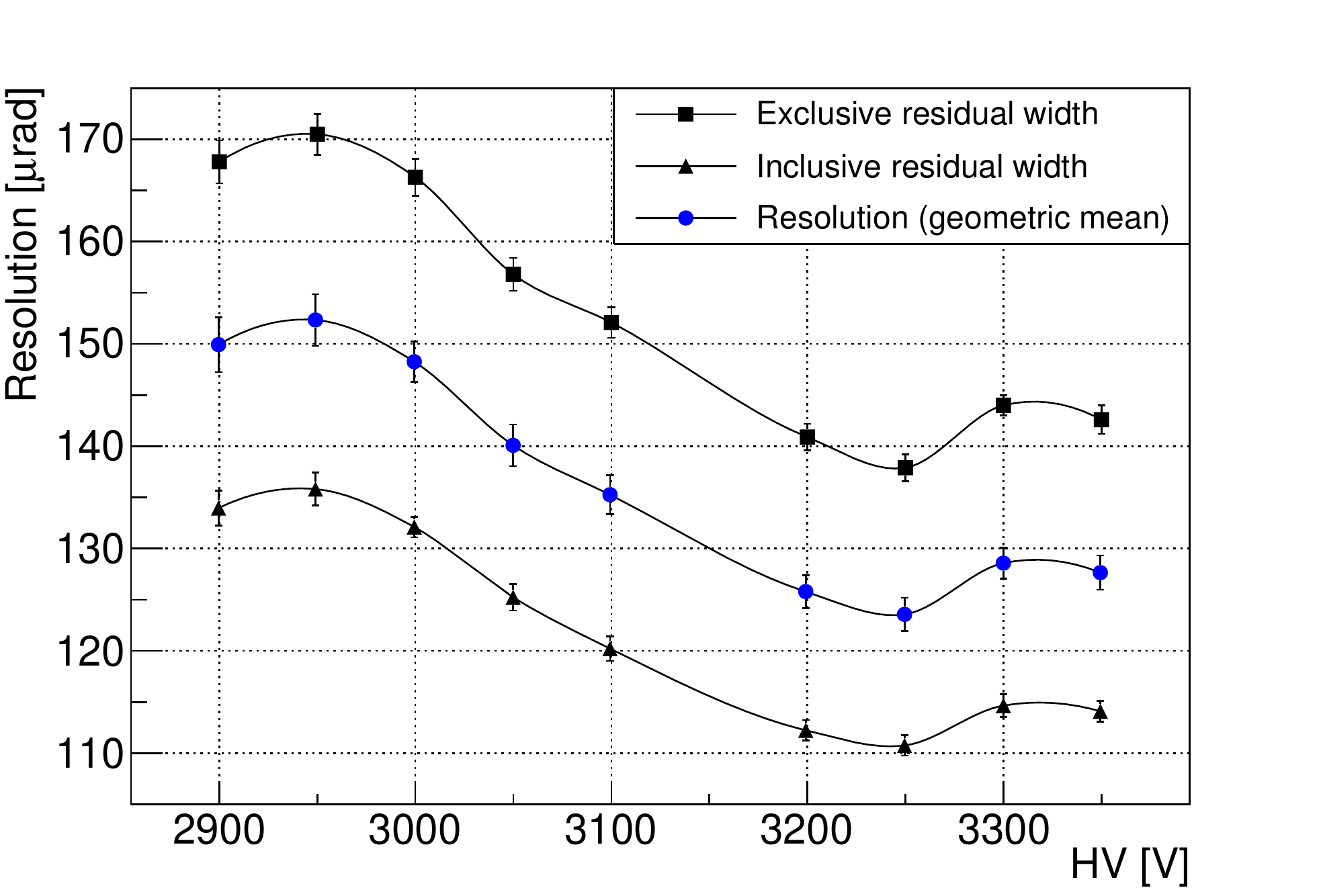}
\caption{Resolutions of the GE1/1 detector at the center of sector 5 for different voltages using full pulse height information.}
\label{label22}
\end{figure}
Residuals for sector 5 are plotted against the different applied high voltages in Fig. 17. The best resolution is obtained on the efficiency plateau as expected. \\

\subsubsection{The Binary Method (Emulated VFAT Threshold)}
During this beam test, all data were collected using APV25 hybrid chips that provide the analog signal, whereas for the CMS upgrade electronics the VFAT3 is proposed that produces a binary output for each readout strip (charge above or below threshold). Hence, it is important to study the characteristics of the GE1/1 detector using binary output. By applying a fixed threshold cut on the pedestal width, binary hits are reconstructed offline from the pulse height data. The detector efficiency is calculated again by applying fixed cuts of \mbox{0.8 fC,} 0.98 fC, and 1.2 fC, which are equivalent to 10 VFAT, \mbox{12 VFAT,} and 15 VFAT units; where 1 VFAT unit = 0.08 fC. The efficiency curve is again fitted with the sigmoid function and from its parameter the efficiency of the detector is found to be [96.9 $\pm$ 0.2 (stat)]\% on the plateau.

\begin{figure}[!htb]
\centering
\includegraphics[width=0.5\textwidth, height=2.3in]{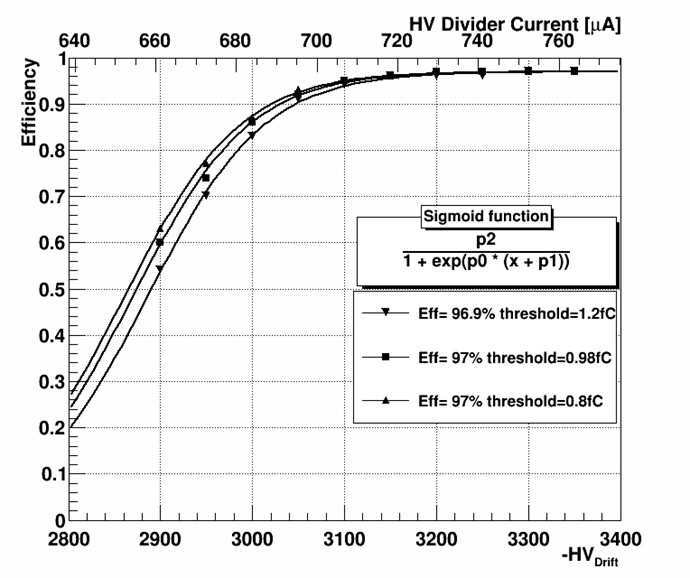}
\caption{Detection efficiency using VFAT-like binary hit data with three different thresholds.}
\label{label15}
\end{figure}

\begin{figure}[!htb]
\centering
\subfigure[]{\label{label15-1}}
\centering
\includegraphics[width=0.5\textwidth]{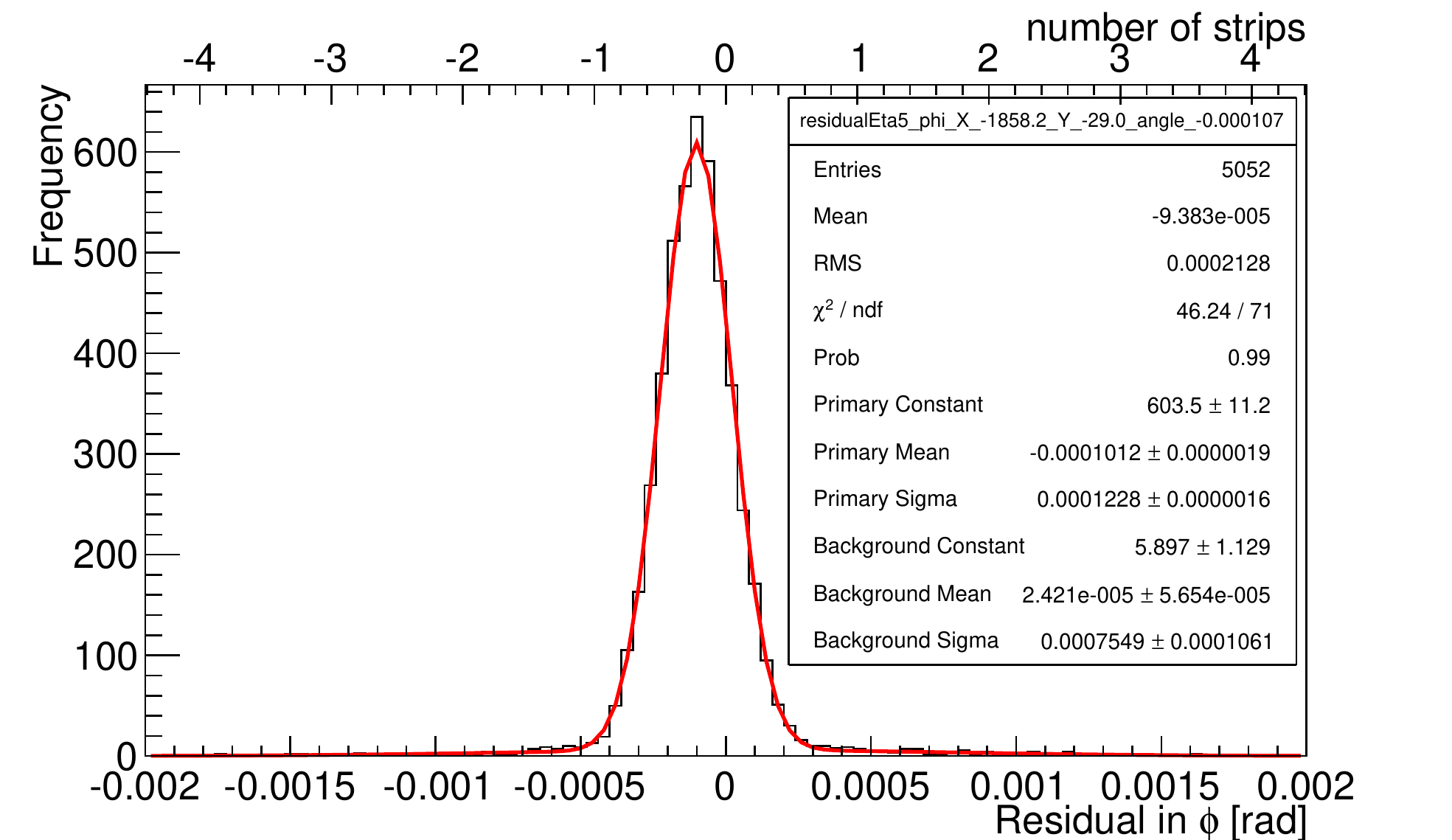}
\subfigure[]{\label{label15-2}}
\centering
\includegraphics[width=0.5\textwidth]{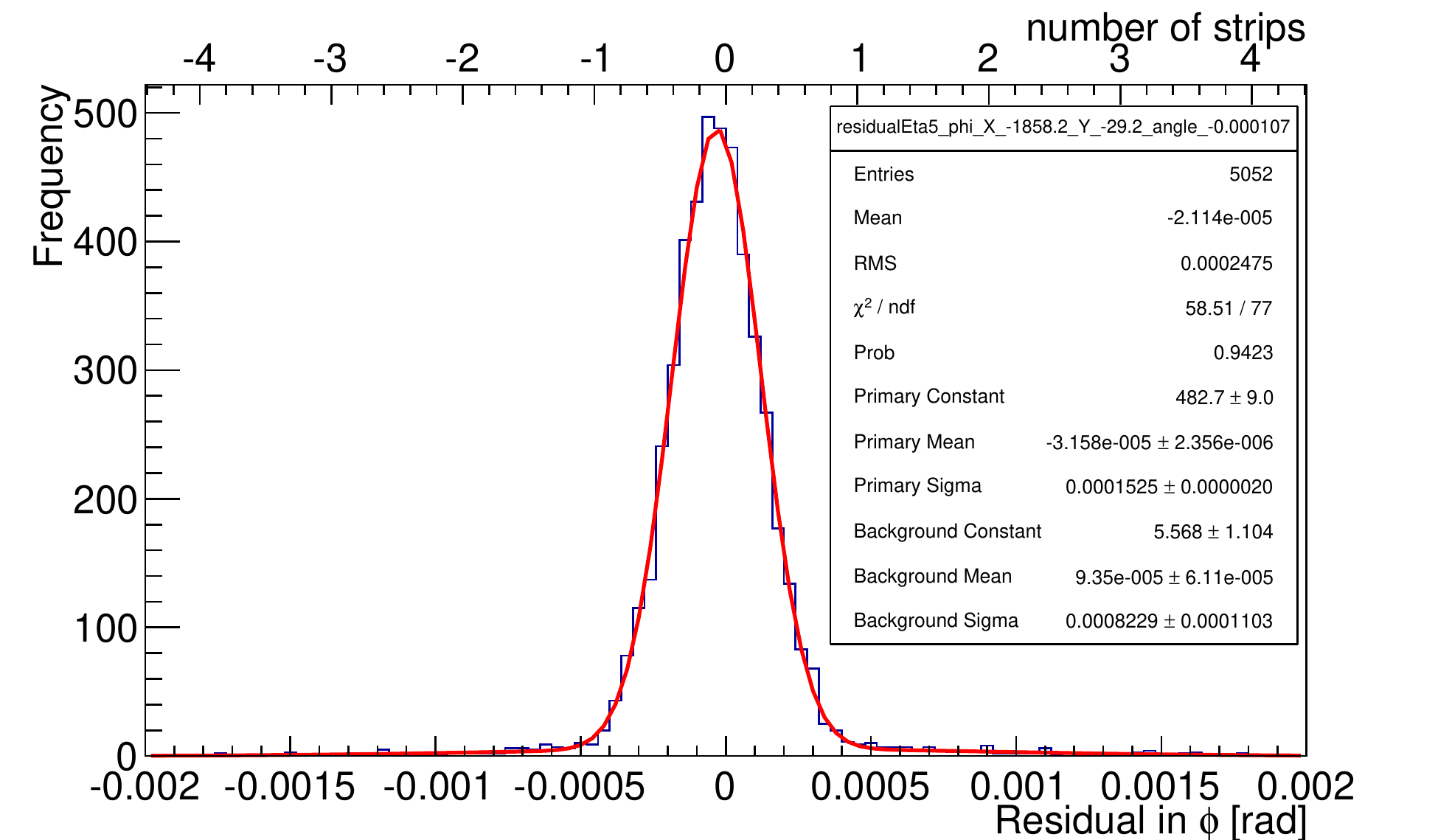}
\caption{(a) Inclusive GE1/1 residual with $\sigma$ = 122.8 $\mu$rad and (b) Exclusive GE1/1 residual with $\sigma$ = 152.5 $\mu$rad using binary hit reconstruction and double-Gaussian fits.}
\label{label16}
\end{figure}
Fig. 19 (a) shows that the inclusive residual width of the detector is 122.8 $\mu$rad (228.1 $\mu$m) and (b) shows that the exclusive residual width is \mbox{152.5 $\mu$rad (283.3 $\mu$m)} with binary hits. The geometric mean of these widths is \mbox{136.8 $\mu$rad (254.2 $\mu$m),} which is consistent with the expectation from angular strip \mbox{pitch/$\sqrt{12}$ = 131.3 $\mu$rad.}

The total radiation length of all chambers used in the test beam setup is about $\sim$14\% which causes multiple scattering of the 32 GeV beam particles on the order of $\sim$147 $\mu$rad RMS of the multiple scattering angle. This value is not negligible as it is within the range of the expected residual widths of the detector. We are currently setting up a Monte Carlo simulation of the full tracking setup including material budgets of all chambers to correct for multiple scattering. We expect that the corrected result for the resolution will be somewhat better than the above preliminary results.

\subsection{Correction of non-linear strip response}
The motivation for implementing this correction for the barycentric method is to further improve the spatial resolution of the GE1/1 detector. In this study, the strip cluster position is reconstructed in the detector via the cluster barycenter, or centroid \cite{Gregorio1}. A $\eta$-correction factor is developed to correct the barycenter position in the GE1/1 detector for biases due to the discretized readout with strips. For the strip cluster with more than one strip, $\eta$ is defined as $\eta = s_{b} -s_{max}$ \cite{Marco}, where $s_{b} = \sum_{i=1}^{n} \frac{q_{i}\cdot s_{i}}{q_{total}}$ gives the barycenter position in terms of strip numbers; $s_{i} $ and $q_{i}$ are strip number and charge of the $i^{th}$ strip, respectively; $s_{max}$ is the strip number with the maximum charge. This correction was done for 2, 3, and 4-strip clusters.
\begin{figure}[!htb]
\centering
\subfigure[]{\label{label18-1}}
\centering
\includegraphics[width=0.24\textwidth, height= 1.25in]{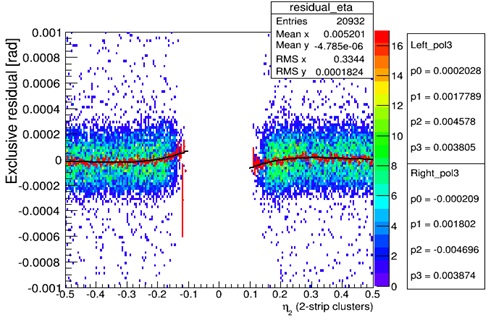}
\subfigure[]{\label{label18-2}}
\centering
\includegraphics[width=0.23\textwidth, height= 1.25in]{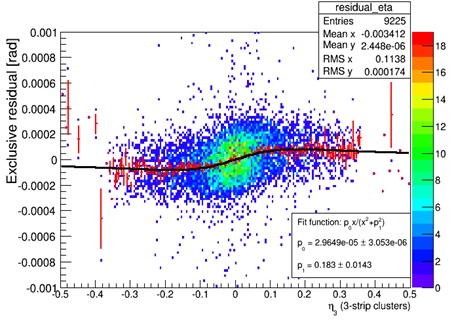}\\
\subfigure[]{\label{label23-1}}
\centering
\includegraphics[width=0.23\textwidth, height= 1.25in]{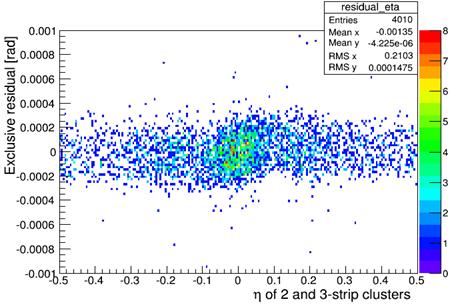}
\subfigure[]{\label{label23-2}}
\centering
\includegraphics[width=0.23\textwidth, height= 1.25in]{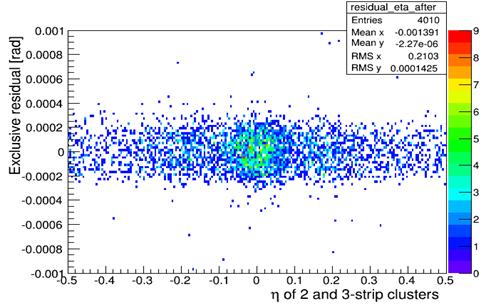} 
\caption{(a) and (b) Scatter plots of exclusive residual vs. $\eta$ for 2-strip and 3-strip cluster, respectively, for combined HV Scan data. (c) and (d) Scatter plots of exclusive residual vs. $\eta$ for all strip cluster multiplicities before and after correction, respectively.}
\label{label18}
\end{figure}
 Fig. 20 (a) and (b) show the scatter plots of exclusive residual against $\eta$ for 2-strip and 3-strip clusters, respectively. They show different behavior and hence are fitted with different functions. Finally, a corrected resolution is obtained for the GE1/1 detector after correcting the measured hit positions by subtracting the value of the fitted function at $\eta$ from the original hit position.
Fig. 20 (c) shows a scatter plot of the residuals for 2-strip and 3-strip clusters before correcting the hit position and (d) shows the corrected residuals for these strip clusters. 
\begin{figure}[!htb]
\centering
\subfigure[]{\label{label19-1}}
\centering
\includegraphics[width=0.53\textwidth, height=1.9in]{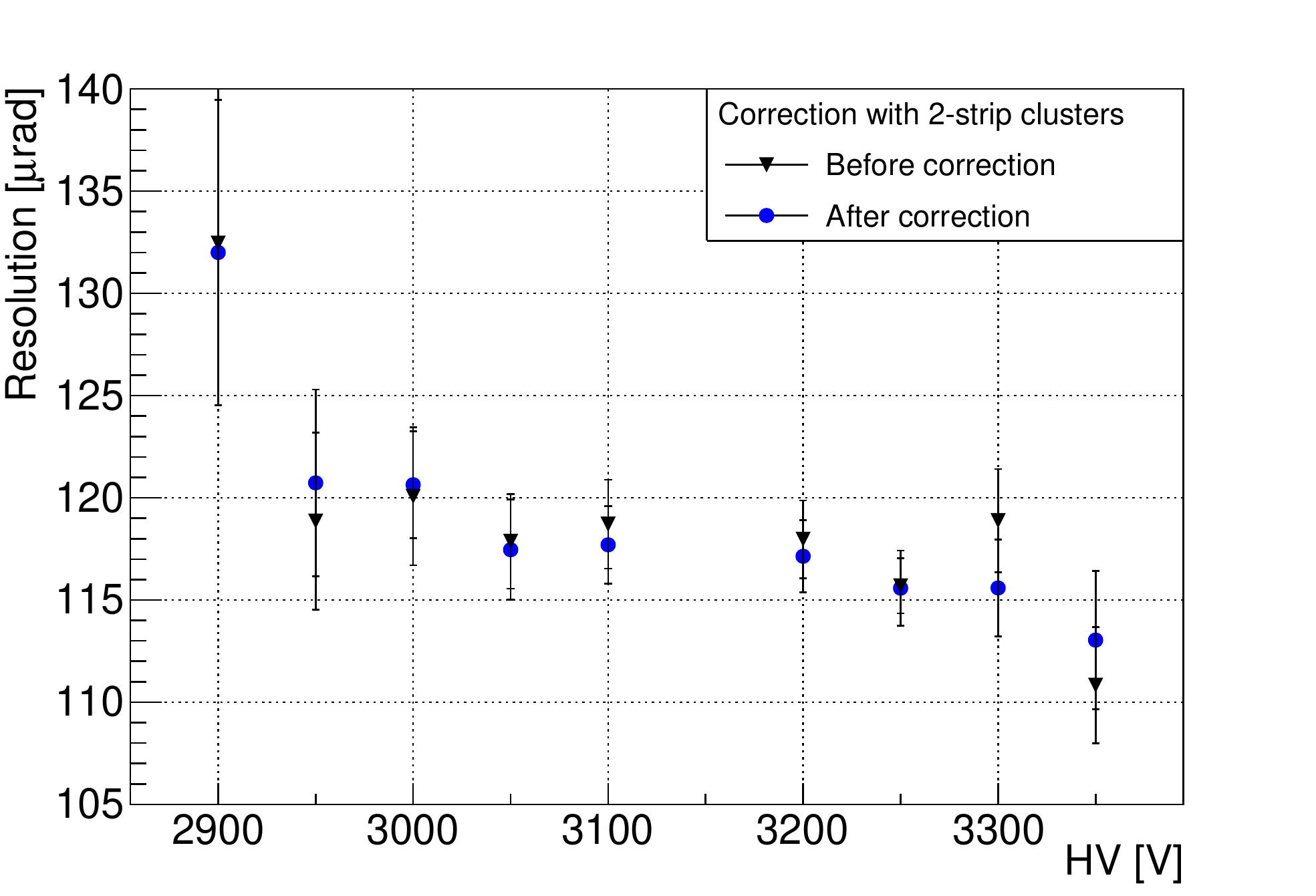}\\
\subfigure[]{\label{label19-2}}
\centering
\includegraphics[width=0.53\textwidth, height=1.9in]{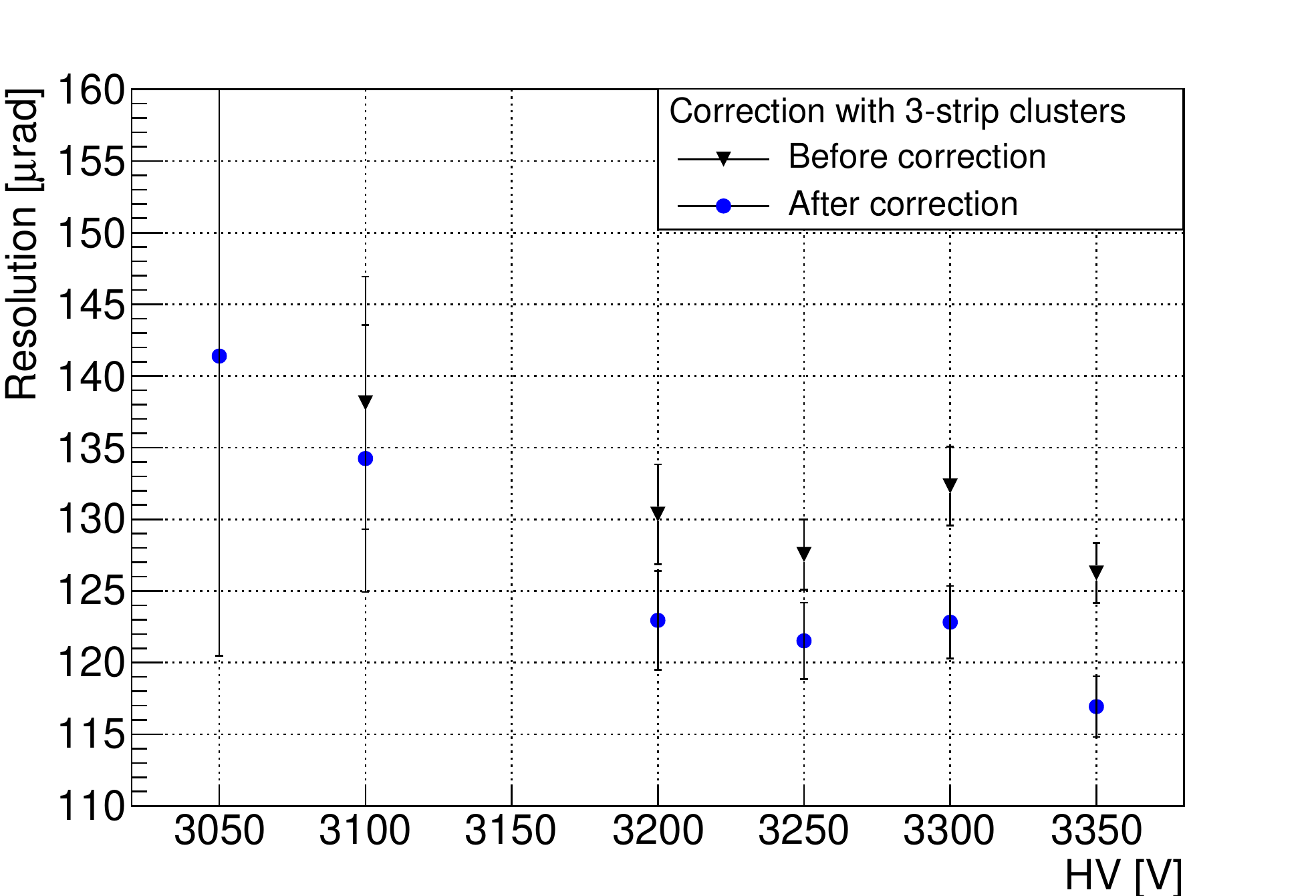}\\
\subfigure[]{\label{label19-3}}
\centering
\includegraphics[width=0.53\textwidth, height=1.9in]{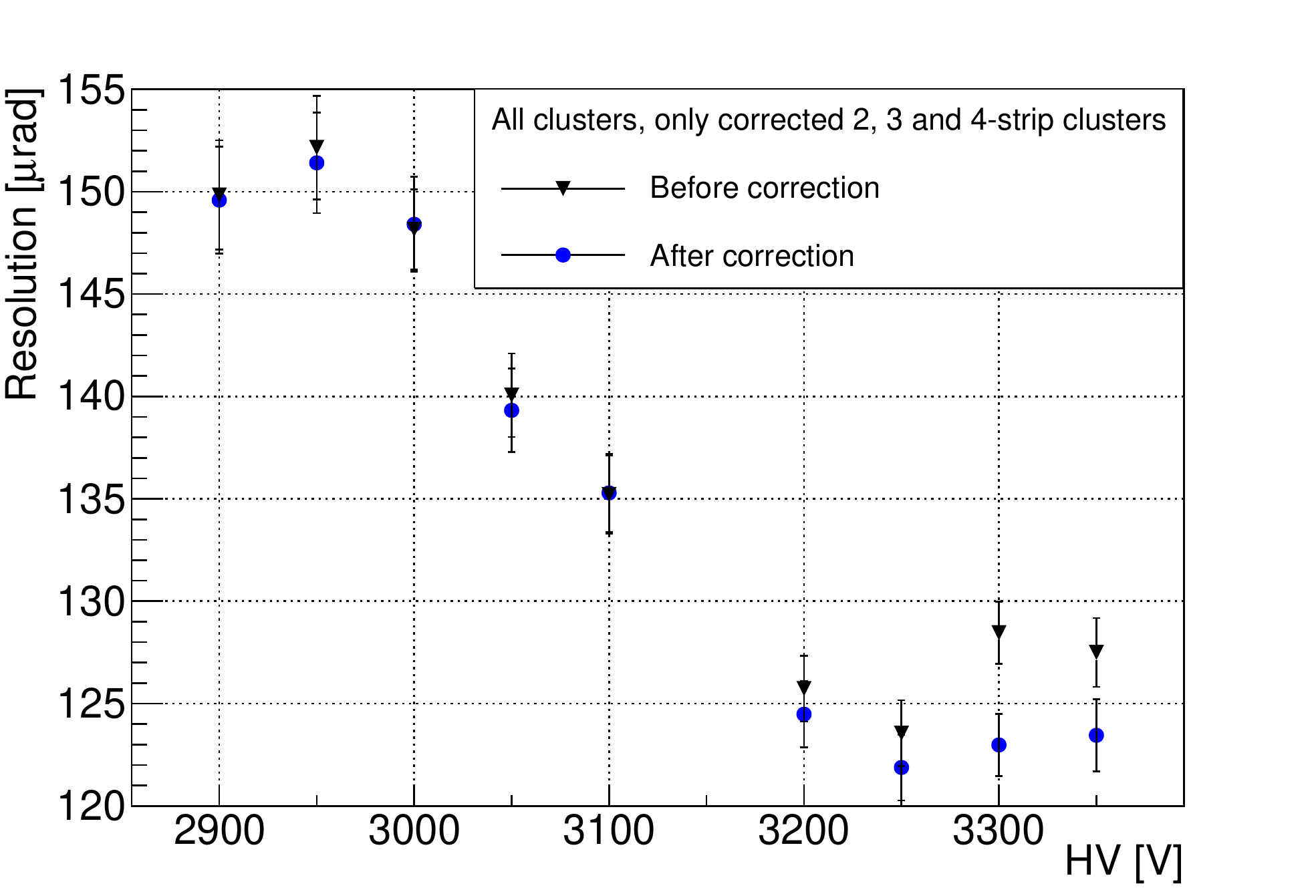}
\caption{(a) Resolutions before and after correction for 2-strip cluster. (b) Resolutions before and after correction for 3-strip cluster. (c) Resolutions before and after correction for 2-strip, 3-strip, and 4-strip cluster for high voltage scan.}
\label{label19}
\end{figure}
For the high voltage scan data, the strip correction factor changes negligibly over the entire voltage range. Consequently, all HV scan data were combined before implementing the strip correction for more statistics. However, the strip correction was performed individually on each sector for the position scan. Fig. 21 shows resolutions before and after correction. Since the readout of the GE1/1  has a fine strip segmentation, the overall improvement factor is small (less than 8\%, i.e. within $\sim$ 10 $\mu$rad) compared to detectors having coarser readout strips \cite{Aiwu}.

\section{Summary and Conclusion}

A CMS GE1/1-III prototype GEM detector was successfully built and tested at Florida Tech and Fermilab, respectively, in 2013. The detector performed well in terms of detector efficiency, which is greater than 97\% with two different methods for defining hits. It shows good charge uniformity for all $\eta$-sectors except for sectors 6 and 7. The uniformity should be improved in future assemblies by making sure that all the foils have the same tension in all eight sectors. The spatial resolution of the GE1/1 detector is $\sim$123 $\mu$rads with the barycentric method and $\sim$136 $\mu$rads with the binary method. The binary method meets the expected value of resolution from the pitch of the strip. The spatial resolution of the barycentric mehod of the detector is improved by $\sim$10 $\mu$rad after correcting for non-linear strip response. In conclusion, the tested GE1/1 prototype detector meets all but one performance expectation for use in the CMS muon endcap upgrade.

\section{Acknowledgement}
We thank the RD51 collaboration for its technical support. Also we thank the FLYSUB consortium as well as the staff of the Fermilab Test Beam Facility for their support during the beam test. Also, we gratefully acknowledge support from FRS-FNRS (Belgium),
FWO-Flanders (Belgium), BSF-MES (Bulgaria), BMBF (Germany), DAE
(India), DST (India), INFN (Italy), Uni. Roma (Italy), NRF (Korea),
QNRF (Qatar), DOE (USA), and WSU (USA).
\newline

\end{document}